\documentclass[a4paper,11pt]{article}
\usepackage[top=25mm,left=22mm]{geometry}
\usepackage{amsmath,amssymb,amsthm,amsfonts}
\usepackage{lineno,url}\usepackage{float}
\floatstyle{plaintop}\restylefloat{table}
\usepackage[tableposition=top]{caption}
\usepackage{graphicx}
\usepackage{multirow} 
\usepackage{pstricks}
\usepackage{fancyvrb}
\usepackage{listings}
\usepackage{paralist, longtable}
\newtheorem{theorem}{Theorem}[section]
\newtheorem{remark}[theorem]{Remark}
\def\pa{{\partial}}\def\lam{\lambda}\def\noi{\noindent}

\def\CO{{\cal O}}\def\CF{{\cal F}}

\newcommand{\barr}{\begin{array}}\newcommand{\earr}{\end{array}}
\newcommand{\bpm}{\begin{pmatrix}}\newcommand{\epm}{\end{pmatrix}}
\def\ri{{\rm i}}\def\ii{{\rm i}}

\def\cc{{\rm c.c.}}\def\er{{\rm e}}\def\ee{{\rm e}}
\def\im{{\rm Im}}
\def\Om{\Omega}
\def\hot{{\rm h.o.t}}
\newcommand{\R}{{\mathbb R}}\newcommand{\C}{{\mathbb C}}

\newcommand{\reff}[1]{(\ref{#1})}\newcommand{\ov}[1]{{\overline {#1}}}
\def\brem{\begin{remark}}\def\erem{\end{remark}}
\newcommand{\bce}{\begin{center}}\newcommand{\ece}{\end{center}}
\newcommand{\bci}{\begin{compactitem}}\newcommand{\eci}{\end{compactitem}}
\newcommand{\bcen}{\begin{compactenum}}\newcommand{\ecen}{\end{compactenum}}
\newcommand{\vs}[1]{{\vspace{#1}}}

\def\eps{\varepsilon}
\def\ra{\rightarrow}
\def\huga#1{\begin{gather} #1 \end{gather}}
\def\hugast#1{\begin{gather*} #1 \end{gather*}}

\def\hualst#1{\begin{align*} #1 \end{align*}}

\def\eex{\hfill\mbox{$\rfloor$}}\DeclareMathOperator{\Res}{Res}
\def\Del{\Delta}\newcommand{\spr}[1]{\left\langle #1 \right\rangle}
\newlength{\tew}

\def\ig{\includegraphics}

\def\pdep{{\tt pde2path}}\def\mlab{{\tt matlab}}

\def\medskip{}\def\bigskip{}

\def\lam{\lambda}\def\ddT{\frac{{\rm d}}{{\rm d}T}}
\def\ass{{\tt ampsys}}
\def\Lh{\hat{L}}\def\uh{\hat{u}} 
\usepackage{hyperref}

\newcommand{\ep}{\varepsilon}
\renewcommand{\im}{{\rm i}}%\renewcommand{\e}{{\rm e}}

\providecommand{\ali}[1]{\begin{align}#1\end{align}}
\providecommand{\alinon}[1]{\begin{align*}#1\end{align*}}

\definecolor{lightgray}{gray}{0.5}

\definecolor{codegreen}{rgb}{0,0.6,0}
\definecolor{codegray}{rgb}{0.5,0.5,0.5}
\definecolor{codepurple}{rgb}{0.58,0,0.82}
\definecolor{backcolour}{rgb}{0.95,0.95,0.92}
 
\lstdefinestyle{mystyle}{
    backgroundcolor=\color{backcolour},   
    commentstyle=\color{codegreen},
    keywordstyle=\color{black},
    numberstyle=\small\color{codegray},
    stringstyle=\color{codepurple},
    basicstyle=\footnotesize\ttfamily,
    breakatwhitespace=false,         
    breaklines=true,                 
    captionpos=b,                    
    keepspaces=true,                 
    numbers=left,                    
    numbersep=5pt,                  
    showspaces=false,                
    showstringspaces=false,
    showtabs=false,                  
    tabsize=2, 
  xleftmargin=4mm,
}
 
\begin{document}
\author{Hannes Uecker\footnote{Institut f\"ur Mathematik, Universit\"at Oldenburg, D-26128 Oldenburg, Germany; hannes.uecker@uol.de},\ \  
Daniel Wetzel\footnote{danieldwetzel@gmail.com}}
\title{The \ass\ tool of \pdep}
\maketitle
\begin{abstract}
The computation of coefficients of amplitude systems for Turing bifurcations 
is a straightforward but sometimes elaborate task, in particular for 
2D or 3D wave vector lattices. The \mlab\ tool 
\ass\ automates such computations for two classes of problems, 
namely scalar equations of Swift--Hohenberg (SH) %and Kuramoto-Sivasinsky (KS) 
type and generalizations, and reaction--diffusion systems with an arbitrary 
number of components. The tool is designed to require minimal user input, 
and for a number of cases can also deal with symbolic computations. 
After a brief review of the setup of amplitude systems we explain the 
tool by a number of 1D, 2D and 3D examples over various wave vector lattices. 
\end{abstract}
%\tableofcontents

\section{Introduction}
\begin{figure}[ht]
\bce 
\begin{tabular}{lll}
(a)&(b)&(c)\\
\ig[width=0.3\textwidth]{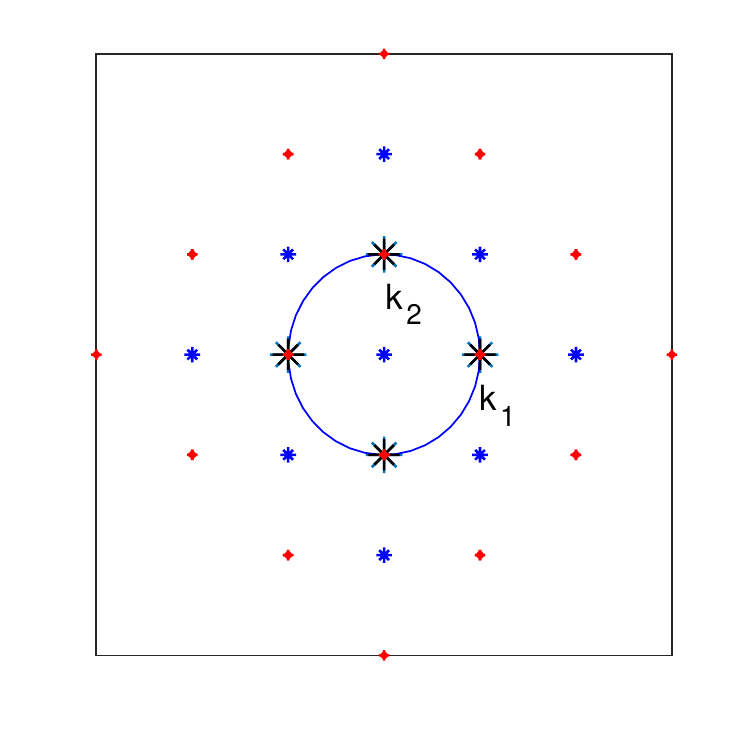}
&\ig[width=0.3\textwidth]{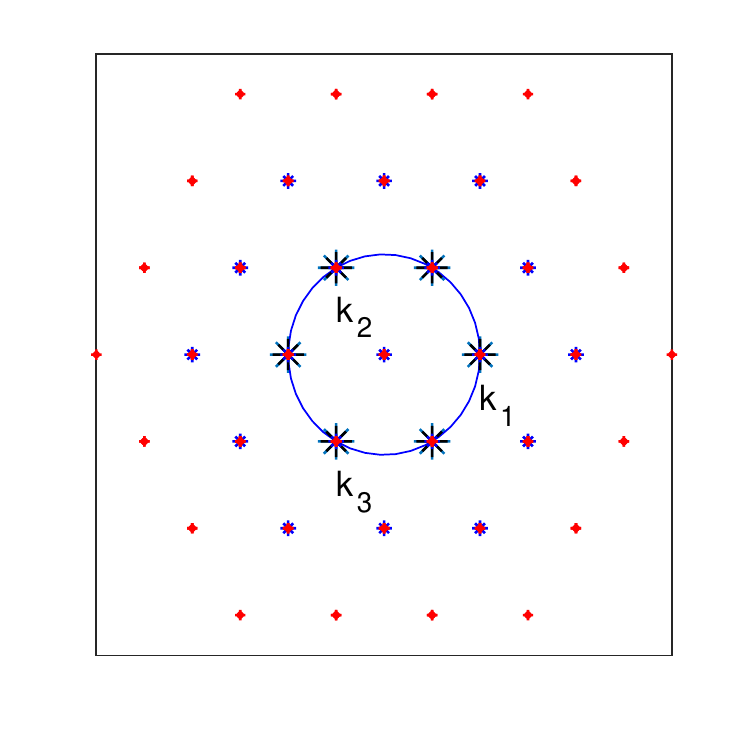}
&\ig[width=0.3\textwidth]{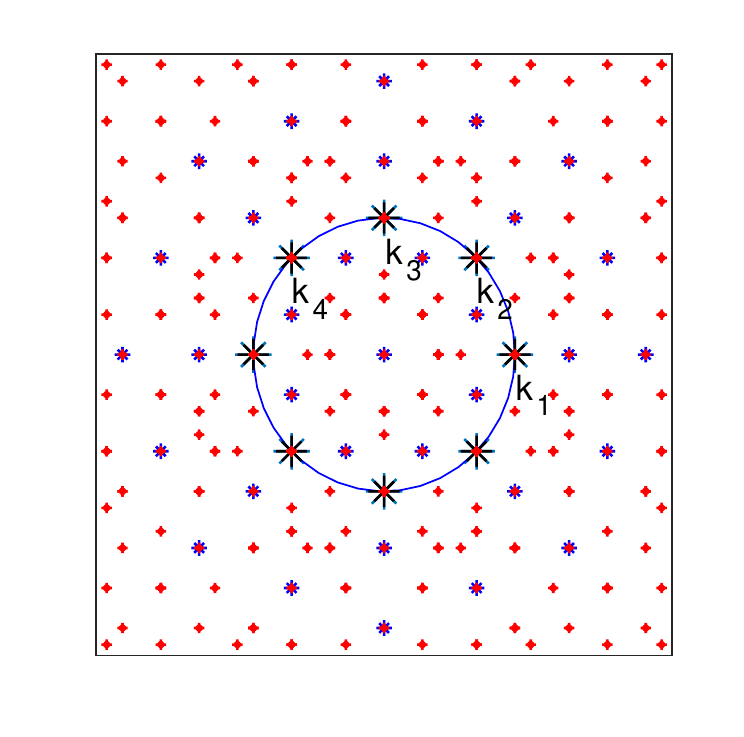}
\end{tabular}
\ece
\vs{-2mm}
   \caption{{\small First three layers (wave vectors generated by terms up to 
cubic order) for (a) Square lattice; (b) hexagonal lattice; (c) 
8-fold quasilattice. 
The thick black stars are the basic wave vectors (on the critical circle), 
the blue (red) dots 
are generated by quadratic (cubic) interactions. 
  \label{lfig}}}
\end{figure}

The Turing bifurcation in a pattern forming system
close to onset is usually described by (systems of) amplitude equations (AEs),  
also called Landau equations. These are ODEs for the amplitudes of
the critical modes, and their derivation, based on center--manifold reduction
or Liapunov-Schmidt reduction, is essentially
a mechanical task, but may become elaborate if the bifurcation is of higher
multiplicity, e.g., due to symmetries of the domain in higher space dimensions.
Such symmetries and the associated AEs have been
classified and analyzed
in detail, see \cite{GoS2002,hoyle} and the references therein.
Typical examples include, e.g., wave-vector lattices of square and hexagonal
type in two space dimensions (2D), see Fig.~\ref{lfig}, 
and simple cubes (SCs), face centered cubes
(FCCs) and body centered cubes (BCCs) in 3D. 

Given a bifurcation problem as above, the tool \ass, included in 
\pdep\ \cite{p2phome}, can be used to compute
the coefficients of the AEs with minimal user input. 
We proceed by example, and illustrate the usage of \ass\ 
to compute the AEs for Swift--Hohenberg (SH) type scalar equations,
and for reaction--diffusion systems (RDS), over 1D, 2D and 3D domains
corresponding to various wave-vector lattices. 
The class of SH type equations is of the form 
\huga{
\pa_t u=Lu+\lam u+c_2 u^2+c_3 u^3, \quad Lu=-(1+\Delta)^2 u \text{\ \ (or similar)}, \label{1cmp}
}
where $u=u(x,t)\in\R$, $t\ge 0$, $x\in \R^d$, where $\lam\in\R$ is the 
bifurcation parameter, and where $c_2,c_3$ can either 
be real coefficients, or operators such as $(\pa_{x_1}+\ldots+\pa_{x_d})u^2$, 
thus including Kuramoto-Sivashinsky (KS) type of equations. 
The RD systems are of the form 
\ali{
u_t=D \Delta u+f(u),\label{rds}
}
where $u\in\R^N$ ($N\ge 2$ components), $D\in\R^{N\times N}$ is a diffusion matrix, and $f:\R^N\ra\R^N$. 

%and here only preview the general setup.
Essentially, the user has to provide (for the RD class) 
\bci  
\item The diffusion matrix $D$ and  a function handle for $f$. 
\item a (spatially homogeneous) steady state $u^*$, and the parameter value(s) 
  where the Turing bifurcation occurs;
\item the critical wave number, and a choice of a wave-vector
  lattice for the amplitudes.
\eci
%The nonlinearity is then internally Taylor-expanded to third order around $u^*$. 
For the SH class, the setup is even simpler and the user has to provide 
\bci
\item the symbol of the linear operator $L$ in Fourier 
space such as $\Lh\uh=-(1-|k|^2) $, and the coefficients $c_2$ and $c_3$.
\eci 
In both cases, this data is to be put into a Matlab struct, for simplicity in the following
called {\tt p} as in {\tt p}roblem. Then calling {\tt [Q,C,c1,phi]=ampsys(p)} returns
\bci
\item the coefficients (or selected coefficients) up to third order in the AEs for the given lattice, 
and (optionally) the critical eigenvector.
\eci
%For the case of standard lattices of $m\ge 2$ wave vectors, the 
%coefficients in the AEs for $\ddT A_j$, $j=2,\ldots,m$ are usually 
%easily obtained from symmetry, but for simplicity (and checking), 
%the user can also request the coefficients in the other equations 
%explicitly. 

\paragraph{Setup:} \ass\ is included in the numerical continuation and bifurcation
  package \pdep \cite{p2pure, p2phome}, with the demo files in
  {\tt demos/asdemos}. In this case, the \mlab\ path to \ass\ is
  already set together with the path to all other \pdep\ library functions
  via {\tt setpde2path} \cite{qsrc}. However, \ass can also be downloaded
  from \cite{p2phome} as a standalone tool, in which case the path is
  set by calling {\tt setastool} in the \ass\ root directory, with the 
demos in the subdirectory {\tt asdemos}. \\[2mm]

\brem\label{lattrem}{\rm Panels (a) and (b) of Fig.~\ref{lfig} show 
the  square and hexagonal lattices (first three 'layers'). For 
such lattice problems, the justification of the amplitude equations 
we compute follows from center--manifold reduction. (c) illustrates 
the quasi-lattice generated by $\vec{k}=\pm\er^{\ri j\pi/4}, j=0,\ldots,3$. 
For this,the nonlinearity generates wave vectors arbitrary close 
to the critical circle (the full lattice is dense in $\R^2$), 
and thus the justification of amplitude 
equations (or the right truncation order) 
becomes a small divisor-problem. See, e.g., \cite{IR2010} and 
the references therein. In \ass, we compute the amplitude equations to 
third order, and thus the distinction between lattices and quasi-lattices 
plays no role.}\eex\erem 

\noi
   {\bf Acknowledgment.}  The work of DW was supported by the DFG under Grant No.~264671738. 

\section{Some amplitude systems on simple lattices as
  analytic examples}\label{asec}
We briefly review the amplitude formalism (AF);
users familiar with Turing bifurcations and the AF, and mainly interested 
in the setup and use of \ass, can safely skip 
this section. Moreover, we restrict to just the formal derivation 
of the AEs, and except for a few remarks 
refer to \cite{hoyle,SU17,pftut} and the references therein 
for justification and 
further conclusions, for instance on special solutions of the AEs. 
See also, e.g., \cite{uwsnak14, w18, UW18b} for comparisons of the 
solutions constructed via the AF with (numerical) solutions of the 
associated full pattern forming systems. 

\subsection{The quadratic-cubic Swift--Hohenberg equation}\label{sha}
Consider the (quadratic-cubic) Swift-Hohenberg (SH) equation 
\begin{equation} \label{swiho} \pa_t u = -(1+\Delta)^2 u + \lam u  +c_1u^2+c_3u^3, 
\quad u=u(x,t) \in \R,\ x\in\Om, 
\end{equation}
where $\Om\subset \R^d$, $d=1,2,3$ (1D, 2D and 3D case, 
respectively), 
with instability parameter $\lam\in\R$,  nonlinearity parameters $c_{2,3}
\in\R$, 
and, if $\Om\ne \R^d$, boundary conditions (BC), 
for instance of the form $\pa_n u|_{\pa\Om}=\pa_n (\Delta u)|_{\pa\Om}=0$.
The original (cubic) SH model \cite{sh77} corresponds to $c_2=0$ and $c_3=-1$, 
while the case $f(u)=\tilde{c}_3 u^3-u^5$ instead of $f(u)=c_2 u^2-u^3$ 
is called the cubic-quintic SH equation. Swift--Hohenberg equations of this 
type are canonical and much studied model 
problems for pattern formation in dissipative systems 
\cite{CH93, pismen06, SU17}.

For all $\lam\in\R$,  \reff{swiho} has
the spatially homogeneous state $u^* \equiv 0$ (trivial branch). 
For $\Om=\R^d$, the linearization $\pa_t v=-(1+\Delta)^2v+\lam v$  
at $u^* \equiv 0$ has the solutions $v(x,t)
= \ee^{ \ii k\cdot x+ \mu(k) t }$, $k\in\R^d$, where
\begin{equation}\label{dr1}
  \mu(k, \lam) =- (1-|k|^2)^2+  \lam, \quad |k|^2:=k_1^2+\ldots+k_d^2. 
\end{equation}
Thus, $u^*\equiv 0$ is asymptotically stable for $\lam<0$, and unstable 
for $\lam>0$ with respect to periodic waves with wave vector $k$
with $|k|=k_c = 1 $. 
%and for suitable domains and BCs we expect bifurcations of spatially $2\pi$ periodic patterns at $\lam=0$. 

\brem\label{qcrem1}{\rm 
Besides the 'classical' SH equation \reff{swiho}, with dispersion 
relation \reff{dr1}, we can also consider equations for which the 
linearization shows simultaneous instabilities at different $|k|$, e.g., 
\huga{\label{8th}
Lu=-(1+\Delta)^2(1+q^{-2}\Delta)^2 u+\lam u, 
}
where wlog $q>1$, with dispersion relation 
\begin{equation}\label{dr2}
  \mu(k, \lam) =- (1-|k|^2)^2(1-q^{-2}|k|^2)^2+  \lam. 
\end{equation}
Similar problems are often used as toy models for quasicrystals, 
cf., e.g., \cite{SAKM16}, and, in 2D and 3D, allow 
multitudes of interesting wave vector interactions and associated (quasi) 
patterns. 
In 1D, we essentially only have to distinguish the cases $q\in\{2,3\}$ 
(resonant case) or $q\not\in\{2,3\}$. We come back to this in \S\ref{eshd}.
\eex}\erem 

% \begin{figure}[ht]
% \bce 
% \begin{tabular}{ll}
% (a) &(b)\\
% \raisebox{30mm}{\begin{tabular}{l}
% \ig[width=0.3\textwidth]{dr1}\\
% \ig[width=0.3\textwidth]{dr2}
% \end{tabular}}
% &\ig[width=0.45\textwidth]{dr3}
% \end{tabular}
% \ece
% \vs{-2mm}
%    \caption{{\small (a) Dispersion relations \reff{dr1} (top), and 
% \reff{dr2} with $\kap=2$ (bottom). (b) Dispersion relation for the 
% extended Brusselator, see \S\ref{ebrusd}, as a typical example 
% for a RD dispersion relation. 
%   \label{drfig}}}
% \end{figure}

\subsubsection{1D} \label{sha1d}
For solutions of \reff{swiho} in 1D we make the ansatz
\huga{\label{ba2}
u(t,x)=\eps\Psi_A(t,x):=\eps A_1(T)e_1+\eps^2\left[\frac 1 2 A_0(T)e_0+A_2(T)e_2\right]+\cc+\hot, \quad e_j=\er^{\ri j x}, 
}
where the amplitude scaling $\eps$ is introduced as $\eps^2=|\lam-\lam_c|=
|\lam|$ (since $\lam_c=0$), such that $\eps^2$ is the distance from 
criticality. In general (with $\lam$ a generic name for a bifurcation 
parameter) we shall use the expansion 
\huga{\label{c0def}
\mu(\lam)=\mu'(\lam_c)(\lam-\lam_c)+\CO(|\lam-\lam_c|^2)=:c_1(\lam-\lam_c)+\CO(|\lam-\lam_c|^2)
}
for the critical eigenvalue. For \reff{swiho} this just gives $c_1=1$. 
The amplitudes $A_j=A_j(T)\in\C$ in \reff{ba2} depend on the 
slow time $T=\eps^2 t$, 
$\cc$ stands for the complex conjugate 
of the preceding terms to obtain real valued $u$, and 
$\hot$ denotes higher order terms which are not relevant for the present computation. 
The $\cc$ of, e.g., $A_1e_1$ is also conveniently written as $A_{-1}e_{-1}$. 
% Neumann BC enforce $\Im(A_j)=0$ such that $A_{-j}=A_j$ and for instance 
% $|A_j|^2=A_j^2$, but for the sake of generality we pretend that the $A_j$ are genuinely complex 
% for $j\ne 0$, which, e.g., is the case for periodic BC or homogeneous 
% Dirichlet BC. 

The {\em residual} of a given ansatz is defined as $\Res(u)=-\pa_t u+Lu+f(u)$, 
and the goal is to choose the ansatz such that the residual formally becomes small. 
Plugging \reff{ba2} into \reff{swiho} we first obtain the $\CO(\eps^2)$ terms 
\hugast{
\Res(u)=\eps^2\left(-A_0e_0-9A_2e_2+c_2(2|A_1|^2 e_0+2A_2^2e_2)+\cc\right)
+\CO(\eps^3). 
}
Importantly, we can solve at the modes $e_0$ and $e_2$ to obtain 
\huga{\label{a02} 
\text{$A_0=2c_2|A_1|^2$ and $A_2=\frac 1 9 c_2A_1^2$.}  
}
Then collecting terms at $\CO(\eps^3e_1)$ yields the AE 
\huga{\label{cGL1} 
\ddT A_1=A_1(c_1\lam\eps^{-2}+c_{31}|A_1|^2)\text{ with } c_1=\mu'(0)=1\text{ and } c_{31}=3c_3+\frac{38}9c_2^2.
}
%If wlog we let $c_3=-1$, then 
\reff{cGL1} predicts the bifurcation of 'stripes'
$|A_1|=\sqrt{\lam/c_{31}}$, where $\lam<0$ for $c_{31}<0$ 
(subcritical bifurcation), 
or $\lam>0$ ($c_{31}>0$, supercritical case). The phase of $A_1$ is free,
and determined by the BC for \reff{swiho}. After deriving \reff{cGL1} 
we can either simply set $\eps=1$, or rescale $\lam=\eps^2\tilde{\lam}$, 
to obtain an amplitude equation independent of $\eps$. 

An $\eps$--scaling as in \reff{ba2} is not always possible in a 
consistent way (see below), but if it is, then it is useful 
as the subsequent derivation of expressions for $A_0, A_2$ and 
$\ddT A_1$ is just a matter of sorting wrt the modes $e_j$ and 
powers of $\eps$. 
If we  omit the $\eps$ scaling, then we essentially need to sort 
wrt the modes $e_j$ and powers of $A_1$, but identification of ``equal powers'' 
may be somewhat ambiguous, see below. 

% \brem{\rm Recall that $\mu(\lam)=\lam=\eps^2\mu$, i.e., the critical 
% eigenvalue $\mu$  is directly given by the bifurcation parameter $\lam$, 
% or, $\pa_\lam \mu(\lam)=1$. 
% Moreover, we scaled the amplitude with $\eps$ and hence the 
% distance (in $\lam$) from criticality by $\eps^2$. Thus, the expansion 
% becomes a formally consistent expansion in powers of $\eps$. However, this 
% is not possible in case of so called quadratic wave vector resonances, 
% see below. 
% }\erem 

\subsubsection{2D} \label{sha2d}
In 2D, the most prominent wave vector lattices are (cf.~Fig.~\ref{lfig}), 
\bci
\item squares, given by, for instance 
the two wave vectors $k_1=(1,0)$ and $k_2=(0,1)$;
\item  hexagons, given by, e.g., $k_1=(1,0)$, $k_2=\frac 1 2(-1,\sqrt{3})$ 
and $k_3=\frac 1 2 (-1/2,-\sqrt{3})$.
\eci 
The crucial difference between the two is that for squares the quadratic 
interaction of critical modes only gives stable modes, i.e., modes 
off the critical circle $|k|=1$ such that quadratic terms can be removed 
from the residual as in \reff{a02}. The hexagon lattice supports 
{\em quadratic resonances}, e.g., $k_1=-k_2-k_3$. As a consequence, 
quadratic terms can in general not be removed from the residual, 
but must be kept in the amplitude equations. The same distinction 
will appear in 3D between, e.g., simple cube (SC) lattices and 
body centered cube (BCC) lattices.

\paragraph{Squares.}   We let $e_{m,n}=\er^{\ri(mx+ny)}$, 
and make the ansatz 
\huga{
u=\eps(A_1 e_{1,0}+A_2e_{0,1})+\eps^2\left[\frac 1 2 A_0+A_{1,1}e_{1,1}+A_{-1,1}e_{-1,1}+A_{2,0}e_{2,0}+A_{0,2}e_{0,2}\right]+\cc+\hot, 
\label{shcm1}
}
where again $\cc$ stands for the complex conjugate since we look for real 
solutions. 

 Collecting terms at $\CO(\eps^2)$ and solving for 
$A_0, A_{2,0}, A_{0,2}, A_{1,1}$ and $A_{-1,1}$ yields 
\hualst{
&A_0=2c_2(|A_1|^2{+}|A_2|^2),\quad A_{2,0}=\frac {c_2} 9 |A_1|^2, \quad 
A_{0,2}=\frac {c_2} 9 |A_2|^2, \quad A_{1,1}=2c_2 A_1A_2,\quad 
A_{-1,1}{=}2c_2 A_{-1}A_2, 
} 
and the complex conjugate equations for $A_{-2,0},\ldots,A_{1,-1}$. 
Now collecting terms at $\CO(\eps^3 e_{1,0})$ and $\CO(\eps^3 e_{0,1})$ yields 
the amplitude equations 
\huga{\label{sqae} \ddT\bpm A_1\\ A_2\epm
=\bpm A_1(c_1\lam\eps^{-2}{+}c_{31} |A_1|^2{+}c_{32}|A_2|^2)\\
A_2(c_1\lam\eps^{-2}{+}c_{31}|A_2|^2{+}c_{32}|A_1|^2)\epm, \ c_1=1, \ 
c_{31}=3c_3{+}\frac{38}9c_2^2, \ c_{32}=6c_3{+}12c_2^2. 
}
Note that the nonlinear coefficients in the second equation 
are obtained by flipping $A_1$ and $A_2$, while the 
linear coefficient $c_1$ is the same for all modes, due to the 
rotational invariance of $\Del$. 

\paragraph{Hexagons.} On a hexagonal grid, a natural ansatz is 
\huga{\label{hexa} 
u(x,t)=A_1(t)e_1+A_2(t)e_2(t)+A_3(t)e_3+\cc+\hot, 
}
where $e_j=\er^{\ri k_j\cdot x}$, but where the $\eps$--scaling 
used before is omitted. The reason is that a consistent $\eps$--scaling 
leading to AEs at third order %(in $\eps$ with $\eps^2=|\lam|$) 
is only possible if the quadratic terms ($c_2 u^2$ in \reff{swiho}) are 
small, i.e., $c_2=\CO(\eps)$. Plugging \reff{hexa} into \reff{swiho} 
we obtain for instance the term $2c_2\ov{A_2}\,\ov{A_3}$ at $e_1$, 
i.e., in the equation for $A_1$, and since $L(|k_c|)$ is not invertible 
we can no longer remove it. Thus, we need to keep it, and altogether 
the amplitude system to third order reads 
\huga{\label{hexeq} 
\begin{split}
\dot A_1&=c_1\lam A_1+c_{21}\ov{A_2A_3}+c_{31} |A_1|^2A_1+c_{32}(|A_2|^2+|A_3|^2)A_1,\\
\dot A_2&=c_1\lam A_2+c_{21}\ov{A_1A_3}+c_{31}|A_2|^2A_3+c_{32}(|A_1|^2+|A_3|^2)A_2,\\
\dot A_3&=c_1\lam A_3+c_{21}\ov{A_1A_2}+c_{31} |A_3|^2A_3+c_{32}(|A_1|^2+|A_2|^2)A_3, 
\end{split}
}
where 
\huga{\label{hexcoeff} 
c_1=1, \quad c_{21}=2c_2+\CO(|\lam c_2|), 
\quad c_{31}=3c_3+\CO(|\lam|+c_2^2),\quad 
c_{32}=6c_3+\CO(|\lam|+c_2^2).
} 
In \reff{hexcoeff}, $\lam$ is not scaled, but 
we treat $|\lam|$ and in particular $c_2$ as small, 
to have a {\em consistent} expansion. However, if we just keep all terms 
up to third order in $A_j$, then 
\huga{\label{hexcoeffnc} 
c_{31}=3c_3+\frac{38}{9}c_2^2,\quad 
c_{32}=6c_3+9c_2^2.
} 
In \S\ref{dsec} we recover the formulas \reff{sqae}, and \reff{hexcoeff}, 
and the {\em inconsistent} version \reff{hexcoeffnc}, 
and their 3D analogs, with \ass. For this we provide the switch {\tt p.cons}(istency), where {\tt p.cons=0} (default setting) corresponds to the case 
\reff{hexcoeffnc}, while {\tt cons=1} yields $c_{31}=3c_3$ and $c_{32}=6c_3$. 

\subsection{The Brusselator}\label{bra}
As an example of an RD system we 
consider the Brusselator \cite{brus}
\begin{equation}\label{bruss1}
\begin{aligned}
\pa_t{u}&=a-(b+1)u+u^2v+D_1\Delta u,\\
\pa_t{v}&=bu-u^2v+D_2 \Delta v, 
\end{aligned}
\end{equation}
where $u=u(x,t)$ and $v=v(x,t)$ correspond to chemical concentrations
of an activator and inhibitor, respectively, 
$D_1$ and $D_2$ are their diffusivities, and $A$ and $B$ are parameters. 
%and $\Delta=\pa_{x_1}^2+\ldots+\pa_{x_d}^2$, where $x\in\Om\subset\R^d$. 
Homogeneous steady states are given by $u^*=a$ and $v^*=b/a$. 
We set 
\huga{\label{pchoice} D_1=1\text{ and } D_2=(a/R)^2, 
} 
with $R>0$ a convenient 'unfolding' parameter (see below), 
and consider $b$ as the primary bifurcation parameter. 
$U^*=(u^*,v^*)$ is stable for 
\huga{\label{bcdef}
B<B_c=(1+R)^2,
} 
where a Turing bifurcation occurs with critical wave number $k_c=\sqrt{R/D_1}$. 

Taylor expanding $f=(a-(b+1)u+u^2v,au-u^2v)^T$ at $U^*$ 
to third order and setting
$w=(\tilde{u},\tilde{v})=(u,v)-(u^*,v^*)$ yields 
\begin{align}
  \pa_t  w = L(\Delta) w +B(w,w)+C(w,w,w),\label{dglredu}
\end{align} 
where
$L(\Delta)=J_f
+\left( \begin{smallmatrix} \Delta&0 \\ 0& D_2 
    \Delta \end{smallmatrix} \right)$, with $J_f=J_f(w^*)$ the Jacobian at 
$w^*$, and $B$ and $C$ are symmetric
bilinear and trilinear forms, respectively. For $p,q,r \in \R^2 $ they
have the form 
\begin{align*}
  B(p,q)=\ & \frac{1}{2} f_{uv}(w^*)(p_1q_2+p_2q_1)+\frac{1}{2}(f_{uu}(w^*)p_1 q_1+f_{vv}(w^*)p_2q_2),\\
  C(p,q,r)=\ &\frac{1}{6}(f_{uuu}(w^*)p_1q_1r_1+f_{vvv}(w^*)p_2q_2r_2)\\
           &+\frac{1}{6} \big( f_{uuv}(w^*)(p_1q_1r_2+r_1p_1q_2+q_1r_1p_2)+f_{uvv}(w^*)(p_1q_2r_2+r_1p_2q_2+q_1r_2p_2)\big). 
\end{align*}

As a compromise between overly simple and too elaborate computations, 
we again choose the 2D square lattice $k_1=k_c(1,0)$, $k_2=k_c(0,1)$ of 
critical wave vectors, and make the
ansatz 
\ali{ \label{ansatzschnak} w=\ep\sum_{j=1}^2 A_j
  e_j\phi+\ep\sum_{j=-2}^{-1} A_j e_j\phi, 
} 
where again $\eps>0$ is defined via $\eps^2:=b-b_c$ 
(distance from criticality), 
%The $A_j$ are amplitudes, $x\in \R^2$, 
$e_j=e^{\im k_j \cdot x}$,
$k_1=k_c(1 \ 0)^T$, $k_2=k_c(0 \ 1)^T$, $k_{-1}=-k_1$, $k_{-2}=-k_2$, 
and $\phi$ is the
eigenvector of $\hat{L}(k_c)$ corresponding to the zero-eigenvalue 
$\mu(b_c,k_c)$. 
By default we normalize $\phi$ by $\phi_1=1$.  Letting $I=\{-2.-1,1,2 \}$ 
we obtain 
\ali{\label{qt} B(w,w) = \ep^2 \sum_{i\in I} \sum_{j\in I} A_i A_j e_i
  e_j B(\phi,\phi), 
}  
and to remove the quadratic terms \eqref{qt} from
the residual, we extend the ansatz \eqref{ansatzschnak} to
\ali{\label{newansatzschnak} w=\ep\sum_{j\in I} A_j e_j\phi+\ep^2
  \sum_{i\in I}\sum_{j\in I} d_{ij} A_i A_j e_i e_j.  } We determine
$d_{ij}$ by substituting \eqref{newansatzschnak} into
\eqref{dglredu} and collecting terms at $\mathcal{O}(\ep^2)$, i.e.,
\huga{\label{dij}
\begin{aligned}
  &d_{ij}=-\hat{L}^{-1}(2k_c)B(\phi,\phi) \text{ for } i=j,\\
  &d_{ij}=-\hat{L}^{-1}(0)B(\phi,\phi)\ \ \text{ for } (i,j)\in 
\{(1,-1), (-1,1), (2,-2), (-2,2)\}, \\
  &d_{ij}=-\hat{L}^{-1}(\sqrt{2}k_c)B(\phi,\phi) \text{ for } 
(i,j)\in\{(\pm 1,\pm 2), (\pm 2,\pm 1)\}. 
\end{aligned}
}
To remove terms of order $\ep^3e_i$ from the residual, we extend the
ansatz \eqref{newansatzschnak} to 
\ali{\label{newansatzschnak2}
  w=\ep\sum_{j\in I} A_j e_j\phi+\ep^2 \sum_{i\in I}\sum_{j\in I}
  d_{ij} A_i A_j e_i e_j+\ep^3\sum_{i\in I} \phi_{3i} e_i.  }
Substituting \eqref{newansatzschnak2} into \eqref{dglredu} and sorting
with respect to $\ep^3 e_1, \ \ep^3 e_2, \ \ep^3 e_3, \ \ep^3 e_4$,
yields
\begin{equation}\label{glpr}
  \begin{aligned}
    \ep^3  e_1: \quad -\hat{L}(k_c)\phi_{31}=&-\pa_T A_1{+} \tilde{c}_{1}(b-b_c)\eps^{-2} A_1 {+}\tilde{c}_2 A_1 A_1 A_{-1} {+}\tilde{c}_3 A_1 A_2 A_{-2},\\
    \ep^3   e_2: \quad -\hat{L}(k_c)\phi_{32}=&-\pa_T A_2{+} \tilde{c}_{1}(b-b_c)\eps^{-2} A_2 {+}\tilde{c}_2 A_2 A_1 A_{-1} {+}\tilde{c}_3 A_2 A_2 A_{-2},
% \\    \ep^3     e_{-1}: \quad -\hat{L}(k_c)\phi_{33}=&-\pa_T A_{-1}{+} \tilde{c}_{1}(b-b_c)\eps^{-2} A_{-1} {+}\tilde{c}_2 A_{-1} A_1 A_{-1} {+}\tilde{c}_3 A_{-1} A_2 A_{-2},\\
%     \ep^3   e_{-2}: \quad -\hat{L}(k_c)\phi_{34}=&-\pa_T A_{-2}{+} \tilde{c}_{1}(b-b_c)\eps^{-2} A_{-2} {+}\tilde{c}_2 A_{-2} A_1 A_{-1} {+}\tilde{c}_3 A_{-2} A_2 A_{-2}
  \end{aligned}
\end{equation}
and similar for $A_{-1}, A_{-2}$ with 
\huga{\label{cti}
\begin{aligned}
  \tilde{c}_1&=\pa_b\mu(b_c)\phi,\\
  \tilde{c}_2&=3C(\phi,\phi,\phi){+}2\left(B(\phi,d_{11}){+}B(\phi,d_{1-1}){+}B(\phi,d_{-11})\right)\\
  &= 3C(\phi,\phi,\phi){+}2B(\phi,-\hat{L}^{-1}(2k_c)B(\phi,\phi)){+}4B(\phi,-\hat{L}^{-1}(0)B(\phi,\phi)),\\
  \tilde{c}_3&=6C(\phi,\phi,\phi){+}2\left(B(\phi,d_{2-2}){+}B(\phi,d_{-22}){+}B(\phi,d_{1-2}){+}B(\phi,d_{-21}){+}B(\phi,d_{12}){+}B(\phi,d_{21})\right)\\
  &=6C(\phi,\phi,\phi){+}4B(\phi,-\hat{L}^{-1}(0)B(\phi,\phi)){+}8B(\phi,-\hat{L}^{-1}(\sqrt{2}k_c)B(\phi,\phi)). 
\end{aligned}
} 
By the Fredholm alternative there exists a solution for \eqref{glpr}
if the right hand sides of \eqref{glpr} are in 
ker($\hat{L}(k_c)^H)^\bot$. Thus, let $\psi$ be the adjoint eigenvector of
$\hat{L}(k_c)$ to the zero-eigenvalue, i.e.,
$\hat{L}(k_c)^H \psi=0$, normalized such that
$\langle \phi, \psi \rangle=1$. The scalar products of \eqref{glpr} with
$\psi$ then yield 
\begin{equation}\label{glpr2}
  \begin{aligned}
    \pa_T A_1&= c_1(b-b_c)\eps^{-2}A_1 +c_2 A_1 A_1 A_{-1} +c_3 A_1 A_2 A_{-2},\\
    \pa_T A_2&= c_1(b-b_c)\eps^{-2}A_2 +c_2 A_2 A_2 A_{-2} +c_3 A_2 A_1 A_{-1},
% \\
%     \pa_T A_{-1}&= c_1(b-b_c)\eps^{-2}A_{-1} +c_2 A_{-1} A_1 A_{-1} +c_3 A_{-1} A_2 A_{-2},\\
%     \pa_T A_{-2}&= c_1(b-b_c)\eps^{-2}A_{-2} +c_2 A_{-2} A_2 A_{-2} +c_3 A_{-2} A_1 A_{-1}\\
  \end{aligned}
\end{equation}
with 
\huga{\label{ci}
c_i=\langle\tilde{c}_i,\psi\rangle. 
}
Using 
$A_1=\ov{A}_{-1}$, $A_2=\ov{A}_{-2}$, and returning to unscaled amplitudes 
$A_1,A_2$ and renaming $c_2=c_{31}$, $c_3=c_{32}$, we may write this as 
\begin{equation}\label{tim}
  \begin{aligned}
    \pa_T A_1&= c_1(b-b_c)A_1 +c_{31} A_1 |A_1|^2 +c_{32} A_1 |A_2|^2,\\
    \pa_T A_2&= c_1(b-b_c)A_2 +c_{31} A_2 |A_2|^2 +c_{32} A_2 |A_1|^2.\\
  \end{aligned}
\end{equation}

\brem\label{brem}{\rm 
a) \reff{tim} shows the structure of the amplitude equations, which is completely as in \reff{sqae}, while the coefficients need to be computed from 
\reff{dij}, \reff{cti} and \reff{ci}. This is, essentially, what 
\ass\ does, for any choice of wave vector lattices, including 3D cases. 
For the Brusselator, this has been 
done analytically for a number of lattices. For instance, for 
the square lattice we obtain  
\huga{\label{cijana}
c_1=\frac{a^2}{(1+R)(a^2-R^2)},\quad c_{31}=\frac{-8+38R+5R^2-8R^3}{9R(a^2-R^2)}, \quad c_{32}=2c_{31}, }
b) For the hexagonal lattice, if we substitute the
 ansatz corresponding to \reff{hexa} into $B$, we obtain terms of the form 
\ali{\label{quadterm}
  A_{-2}A_{-3}e_{-2}e_{-3}B(\phi,\phi)=A_{-2}A_{-3}e_1B(\phi,\phi).  
}
Since $\hat{L}(k_c)$ is not invertible, we cannot remove such terms 
from the residual and proceed as in \reff{hexeq}, i.e., keep them. 
We then obtain the amplitude system 
\reff{hexeq} with $\lam=b-b_c$, and, analytically, up to third order, 
but inconsistently in the sense of \reff{hexcoeffnc}, 
$c_1, c_{31}$ as in \reff{cijana}, and 
\cite{VWDB92}
\huga{\label{cij2}
c_{21}=\frac {2a(1+R)(1-R)}{a^2-R^2}, \quad c_{32}=\frac{3-5R+7R^2-5R^3}{a^2 R(1+R)}. 
}
Thus, $c_{21}=0$ for $R=1$, and we should expect \reff{hexeq} to be valid 
only for small $|R-1|$. 
See also \cite{CKnob97, CKnob99} for 3D cases. 
%Here, the formulas \reff{cijana} and \reff{cij2} 
%will again be used to check the implementation of \ass, see below.  
\\
c) In summary,  
given the user data $D$, $u^*$, the parameters for the Turing 
bifurcation, the critical wave number and the choice of wave vector 
lattice, and the function $f$, 
%and data about the active bifurcation parameter $\lam$, 
\ass\ proceeds as follows: % for the RD case: 
\bci
\item Expand $f$ to third order around $u^*$. 
\item  Compute the critical eigenvector $\phi$ (normalized to 
$\phi_1=1$), $c_1=\pa_\lam\mu_c(\lam_c,k_c)$, and the adjoint 
critical eigenvector $\psi$ (normalized to $\spr{\phi,\psi}=1$). 
\item Check if there are quadratic resonances. 
\bci
\item If no, then 
compute the terms $d_{ij}$ as in \reff{dij}, and from these the 
cubic coefficients as in \reff{cti} and \reff{ci} (for more 
complicated lattices, there will be many more coefficients 
to be computed, and to be returned in adequate form, see \S\ref{dsec}). 
\item If yes, then also compute 
(and return) the quadratic coefficients. Moreover, here we need to 
decide if we want a {\em consistent} expansion or not, where as in 
\reff{hexcoeff} consistent means that we do not add quadratic 
corrections to the ansatz for computing the third order terms, 
because the quadratic corrections are (assumed to be) 
 small and hence formally do not show up in the cubic terms. 
 On the other hand, an inconsistent expansion such as \reff{hexcoeffnc} 
 may be more accurate. In \ass, this choice is made by a switch {\tt p.cons}, 
 and the default value $0$ means the inconsistent but more standard 
%and often more accurate 
choice.
\eci
\eci 
For the scalar (SH or KS like) case, the procedure is essentially the same, with $\phi=\psi=1$. The setup to apply this algorithm, and the results, 
are explained in \S\ref{dsec} by a number of examples. 
}\eex\erem 

\section{The demos}\label{dsec}

\def\dhome{./SH}
\subsection{The Swift-Hohenberg equation, demo {\tt SH}}\label{shd}
 As first example we consider \reff{swiho}, and thus in the demo directory {\tt SH} 
set up {\tt L} as in Listing \ref{LSH}. 
Furthermore, \ass\ needs information about $c_2$, $c_3$, and the
wave vectors. In the following we explain this for four cases, 
namely 1D, 2D (square and hex)  and 
a 3D SC. We mainly use symbolic computations, i.e., compute the coefficients 
in the amplitude equations as functions of $c_2,c_3$. 
The script for all demos {\tt SH/cmds.m} is in cell mode, 
which means that the user can and should run cells interactively one-by-one. 

\brem\label{c1rem}{\rm The general calling syntax of \ass\ is 
{\tt [Q,C,c1,phi]=ampsys(p)}, where {\tt p} contains the 
problem description. For the class of SH equations (scalar) in this 
section, always {\tt phi=1}, and for {\tt c1} we just return the 
dummy {\tt c1=0}. The idea is that for SH type equations, 
the user can and should always compute {\tt c1} herself. Thus, for 
SH type equations we can as well just call {\tt [Q,C]=ampsys(p)}. 
}
\erem 

\lstinputlisting[caption={{\small {\tt SH/L.m}, encoding the
    Fourier transform of $L=-(1+\Delta)^2$.}},
label=LSH,language=matlab,stepnumber=5,
firstnumber=1]{\dhome/L.m} 

\begin{lstlisting}[caption={{\small script {\tt SH/cmdsSH.m}, organized in \mlab\ cells, i.e., to be run cell-by-cell.}}, 
label=LSH,language=matlab,stepnumber=5,linerange=1-16, firstnumber=1]
%% C1:  SH 1D, numerical coefficients
p=[]; p.c2=0.1; p.c3=-1; % cleaning p, setting numerical values for parameters
p.k=1; [Q,C]=ampsys(p); % setting kc, and calling ampsys 
%% C2:  SH 1D, symbolic coefficients
p.sb=1; syms c2 c3; p.c2=c2; p.c3=c3;  % symbolic parameters
p.k=1; [Q,C]=ampsys(p); % setting kc, and calling ampsys 
%% C3:  SH 2D, square lattice, symbolic nonlinearity coeff, coeff for both eqns
kc=1; type=21; p.k=wavevec(kc,type); p.eqnr=[1 2]; [Q,C]=ampsys(p); 
%% C4:  SH 2D, hexagon lattice, symbolic nonlinearity coeff, cons=1
kc=1; type=22; p.k=wavevec(kc,type); p.cons=1; p.eqnr=1; [Q,C]=ampsys(p); 
pause; p.eqnr=[2 3]; [Q,C,phi]=ampsys(p); % give coeff for other eqns 
%% C5:  SH 2D, hexagon lattice, symbolic nonlinearity coeff, cons=0
p.cons=0; kc=1; type=22; p.k=wavevec(kc,type); p.eqnr=1; [Q,C]=ampsys(p); 
%% C6:  SH 3D, BCC lattice, symbolic, cons=0; 
p.cons=0; kc=1; type=33; p.k=wavevec(kc,type); p.eqnr=1; [Q,C]=ampsys(p); 
\end{lstlisting}

\subsubsection{1D}\label{shd1d}
\paragraph{Numerical values for parameters.} 
In Cell 1 of {\tt cmdsSH.m} we 
set {\tt p.c2=0.1, p.c3=-1}, and, in 1D,  {\tt p.k=1}, because 
\ass\ automatically uses $-k$ as well, i.e., the ansatz is 
$u=\ep A_1e^{\text{i}x}+\ep A_{-1}e^{-\text{i}x}$. 
The output of {\tt [Q,C]=ampsys(p)} is
\alinon{ {\tt Q}=[],\qquad \text{and} \qquad {\tt
    C}=(1 \ 1 \ -1 \ -2.958). 
} 
The last entry of {\tt C} is the coefficient $c_{31}$, 
while the preceding entries are the indices of the $A_j$, 
which here means that the cubic term in the first equation is 
$-2.958 A_1 A_1 A_{-1}$, i.e., 
\alinon{ \pa_T A_1&=A_1-2.958 A_1 A_1 A_{-1}.
} 
Since $u$ is real, 
$A_{-1}=\ov{A}_1$, and hence $\pa_T A_1=A_1-2.958
  A_1 |A_1|^2$. 

\paragraph{Symbolic parameters.} 
In Cell 2 we switch to symbolic parameters, and obtain, in agreement with 
\reff{cGL1}, ${\tt Q}=[]$ and 
${\tt C}=(1, 1, -1, 3c_3 + 38c_2^2/9)$. 

\subsubsection{2D} \label{shd2d}
\paragraph{Square lattice.} To set the wave-vector lattice
for the square lattice, we can use {\tt p.k=[1,0; 0,1]}. More conveniently, 
we can use the function {\tt p.k=wavevec(kc,type)}, which provides 
the most common lattices, see Table \ref{tabwv}. 
\begin{table}[htpb]
  \centering
  \caption{Using {\tt k=wavevec(kc,type)}} \label{tabwv}
  \begin{tabular}{|c | c | c |}
    \hline
    type & wave vectors & lattice type\\ \hline
    1 & $k={\tt kc}$ & 1D \\
    21 & $k={\tt kc}
         \begin{pmatrix}
           1&0\\ 0 & 1
         \end{pmatrix}$ & square (SQ)\\
    22 & $k={\tt kc}\begin{pmatrix}1&-0.5&-0.5\\
      0&\sqrt{3}/2&-\sqrt{3}/2\end{pmatrix}$ & hexagonal\\
    31 & $k={\tt kc}\begin{pmatrix} 1&0&0\\
      0&1&0\\
      0&0&1
    \end{pmatrix}$ & simple cubic (SC)\\
    32& $k=\frac{ {\tt kc}}{\sqrt{3}}\begin{pmatrix}
      1&1&-1&-1\\
      1&-1&1&-1\\
      1&-1&-1&1\\
    \end{pmatrix}$ & face-centered cubic (FCC)\\
    33 & $ k=\frac{{\tt kc}}{\sqrt{2}}
         \begin{pmatrix}
           1&0&1&1&0&-1\\
           1&1&0&-1&1&0\\
           0&1&1&0&-1&1\\
         \end{pmatrix}
    $
         & body-centered cubic (BCC)\\
    \hline
  \end{tabular}
\end{table}
The output of \ass\ for the squares reads (slightly cleaning up the \mlab\ 
output for {\tt C}) 
$$
\text{{\tt Q=[]} (no quadratic terms) and 
{\tt C=$\bpm 1& 1& -1& 3c_3 + 38c_2^2/9\\
1& 2& -2& 6c_3 + 12c_2^2\epm$}.}
$$ 
Thus, the cubic terms in the first equation 
are $(3c_3+38c_2^2/9) A_1^2 A_{-1}+
(6c_3+12c_2^2)A_1A_2A_{-2})$, i.e., 
\hugast{
\pa_T A_1=c_1A_1+(3c_3+38c_2^2/9) |A_1|^2 A_{1}+
(6c_3+12c_2^2)|A_2|^2A_1, 
}
in agreement with the first equation in \reff{sqae}. To see the coefficients 
in both equations (which follow from symmetry), in C3 
we let {\tt p.eqnr=[1 2]} and 
call {\tt [Q,C]=ampsys(p)}. Then ${\tt Q}=[]$ as before, and 
\hugast{
{\tt C}=\bpm 
1& 1& -1& 38c_2^2/9 + 3c_3& 1& 2& -1&     12c_2^2 + 6c_3\\
1& 2& -2&     12c_2^2 + 6c_3& 2& 2& -2& 38c_2^2/9 + 3c_3
\epm. 
}
The coefficients for $\ddT A_1$ are in the first $2\times 3$ block, 
and those for $\ddT A_2$ in the second $2\times 3$ block, i.e., the second 
equation reads 
\huga{
\pa_T A_2=c_1A_1+(6c_3+12c_2^2)|A_1|^2A_2+
(3c_3+38c_2^2/9) |A_2|^2 A_{2}. 
}

\paragraph{Hexagon lattice.} We let {\tt p.k=wavevec(1,22)} and {\tt p.cons=1} (the consistent choice) and obtain 
$${\tt Q}=(-2, -3, c_2) \text{ and } {\tt C}=\bpm  1&1&-1&3c_3\\
1&2&-2&6c_3\\
1&3&-3&6c_3\epm, 
$$
meaning that  in agreement with \reff{hexcoeff} the quadratic and cubic terms in the first equation 
are $c_2\ov{A_2}\ov{A_3}$ and $3c_3|A_1|^2A_1+6c_3(|A_2|^2+|A_3|^2)A_1$, 
respectively. Note that $c_2$ is considered to be small and hence omitted 
in the expressions in {\tt C}. On the other hand, for {\tt p.cons=0} we recover 
\reff{hexcoeffnc}, i.e., 
$${\tt Q}=(-2, -3, c_2) \text{ and } 
{\tt C}=\bpm  1&1&-1&3c_3+38c_2^2/9\\
1&2&-2&6c_3+9c_2^2\\
1&3&-3&6c_3+9c_2^2\epm.  
$$

\subsubsection{3D BCC} \label{shd3d}
On the BCC lattice, {\tt [Q,C]=ampsys(p)} (with default 
setting {\tt p.cons=0}) yields 
\huga{\label{SHBCC}
{\tt Q}=\bpm 2&-6&2c_2\\3&5&2c_2\epm, \quad 
{\tt C}=\bpm 1&1&-1&3c_3 + 38c_2^2/9\\
1& 2& -2& 6c_3 + 9c_2^2\\
1& 3& -3& 6c_3 + 9c_2^2\\
1& 4& -4& 6c_3 + 12c_2^2\\
1& 5& -5& 6c_3 + 9c_2^2\\
1& 6& -6& 6c_3 + 9c_2^2\\
2& 4& 5 &6c_3 + 12c_2^2\\
3& -4& -6& 6c_3 + 12c_2^2\epm. 
}
Thus, the first amplitude equation is given by 
\huga{\begin{aligned}
\ddT A_1=&\lam A_1+q(A_2\ov{A}_6+A_3A_5)+
 c_{31}|A_1|^2 A_1+c_{32}(|A_2|^2+|A_3|^2+|A_5|^2+|A_6|^2)\\
 &+c_{33}|A_4|^2A_1+c_{34}(A_2A_4A_5+A_3\ov{A}_4\ov{A}_6), 
\end{aligned}
 \label{bccgen}
}
with $q=2c_2$, $c_{31}=3c_3 + 38c_2^2/9$, $c_{32}=6c_3 + 9c_2^2$, 
$c_{33}=6c_3 + 9c_2^2$
and $c_{34}=6c_3 + 12c_2^2$. The form \reff{bccgen} is the {\em general} 
form on a BCC, following from symmetry, which is why we did not 
group together $c_{32}$ and $c_{34}$, which in general are not equal. 

If we enforce consistency via {\tt p.cons=1}, then $c_{31}=3c_3$, 
and always (i.e., independent of the model) 
\huga{\text{$c_{32}=c_{33}=c_{34}=2c_{31}$.} 
}

\subsection{Two length scales pattern formation, demo {\tt qc}}\label{eshd}
\def\dhome{./qc}
To illustrate the flexibility of \ass\ for scalar equations, we consider 
the 8th order equation 
\huga{
\pa_t u=Lu+\lam u+c_2 u^2+c_3 u^3, \quad Lu=-(1+\Delta)^2(1+q^{-2}\Delta) u, 
\label{qce}
}
with dispersion relation $\mu(|k|, \lam) 
=- (1-|k|^2)^2(1-q^{-2}|k|^2)^2+  \lam$. We have $\mu(|k|,0)=0$ 
simultaneously at $|k|=1$ and $|k|=q^2$. Hence, for instance in 1D, 
the ansatz for bifurcating (quasi--)periodic patterns reads 
\huga{\label{qcqp}
u(x,t)=A_1e_1+A_qe_q+\cc+\hot,\quad e_j=\er^{\ri jx},
} 
and in 2D or 3D there are plenty of possibilities to choose wave vectors 
sets describing different (quasi)patterns. Here we restrict to 
amplitude equations for two cases, namely: 1D, and 2D with 
a hexagonal lattice for $|k|=1$ and a square lattice for $|k|=q$. 

For the amplitude equations we must distinguish between 
the resonant case $q\in\{2,3\}$, where, e.g., quadratic ($q=2$) 
or cubic ($q=3$) interactions of $|k|=1$ modes may directly map 
to $|k|=q$ modes, and the non--resonant case $q\not\in\{2,3\}$, 
where such direct couplings do not occur. 
Moreover, the coupling between modes 
belonging to wave vectors with different $|k|$ is no longer symmetric, 
such that, given the coefficients in the first equation, the coefficients 
in the other equations no longer follow from simple symmetries. 
Thus it is useful and convenient to tell {\tt ampsys} 
to compute coefficients for different equations in the amplitude 
system via {\tt p.eqnr}, which, e.g., yields the full system 
via {\tt p.eqnr=1:m} with {\tt m} the number of modes in the lattice. 
Listing \ref{qcl1} 
shows the implementation of $\hat L$ for \reff{qce}, and 
Listing \ref{qcl2} the script file. In any case, recall from Remark 
\ref{lattrem} that the amplitude equations derived here are third 
order truncations of an in general very complicated small divisor problem.

\lstinputlisting[caption={{\small {\tt qc/L.m}, passing the parameter $q$ 
via the problem struct {\tt p}.}},
label=qcl1,language=matlab,stepnumber=5,
firstnumber=1]{\dhome/L.m} 

\begin{lstlisting}[caption={{\small {\tt qc/cmdsqc.m}.}},
label=qcl2,language=matlab,stepnumber=5,linerange=1-11, 
firstnumber=1]
%% C1:  qc 1D, non-resonant case, give coefficients for both eqns 
q=1.5; p.k=[1 q]; p.q=q; % store second critical wave-nr, needed in L 
p.sb=1; syms c2 c3; p.c2=c2; p.c3=c3; p.eqnr=[1 2]; 
[Q,C]=ampsys(p);
%% C2: 1D, resonant case, with cons=1
q=2; p.k=[1 q]; p.q=q; p.cons=1; [Q,C]=ampsys(p);
%% C3: 1D, resonant case, with cons=0
p.cons=0; [Q,C]=ampsys(p);
%% C4: 2D, mix of hex and square lattice, resonant 
p.q=2; p.k=[wavevec(1,22) wavevec(p.q,21)]; 
p.eqnr=[1 4]; [Q,C]=ampsys(p);  % return coeff for first and 4th equation 
\end{lstlisting}

\subsubsection{1D} 
As a warmup example for the non--resonant case let $q=1.5$, see C1. Then 
setting {\tt p.eqnr=[1,2]} and calling 
{\tt [Q,C]=ampsys(p)} yields $Q=[]$, and 
\hugast{
C=\bpm
 1& 1& -1&   214c_2^2/49 + 3c_3& 1& 2& -1& 2557c_2^2/196 + 6c_3\\
 1& 2& -2& 2557c_2^2/196 + 6c_3& 2& 2& -2& 1153c_2^2/288 + 3c_3
\epm. 
}
Hence, the amplitudes $A_1$ and $A_q$ from \reff{qcqp} fulfill 
\hugast{
\barr{rl} \pa_T A_1=&c_1A_1+(c_{31}|A_1|^2+c_{32}|A_2|^2)A_1, \\
\pa_T A_2=&c_1A_1+(c_{32}|A_1|^2+c_{33}|A_2|^2)A_2,\earr
}
with $c_{31}=3c_3 + 214c_2^2/49$, $c_{32}=6c_3 + 2557c_2^2/196$ 
and $c_{33}=3c_3 + 1153c_2^2/288$. 

On the other hand, for, e.g., $q=2$ and {\tt p.cons=1} we obtain 
\hugast{{\tt Q}=(2, -1, 2c_2, 1, 1, c_2), \quad 
{\tt C}=\bpm 
1& 1& -1& 3c_3& 1& 2& -1& 6c_3\\
1& 2& -2& 6c_3& 2& 2& -2& 3c_3
\epm, 
} 
from which we can again write down the system for $\pa_T A_1, \pa_T A_2$. 
For {\tt p.cons=0} we obtain $Q$ as before, and 
\hugast{
{\tt C}=\bpm 
1& 1& -1&        6c_2^2 + 3c_3& 1& 2& -1&    201c_2^2/25 + 6c_3\\
1& 2& -2& 201c_2^2/25 + 6c_3& 2& 2& -2& 8102c_2^2/2025 + 3c_3\\
\epm. 
}
All these formulas can be checked by hand in a straightforward  
although already a bit lengthy way. 

\subsubsection{2D} 
\begin{figure}[ht]
\bce 
\ig[width=0.3\textwidth]{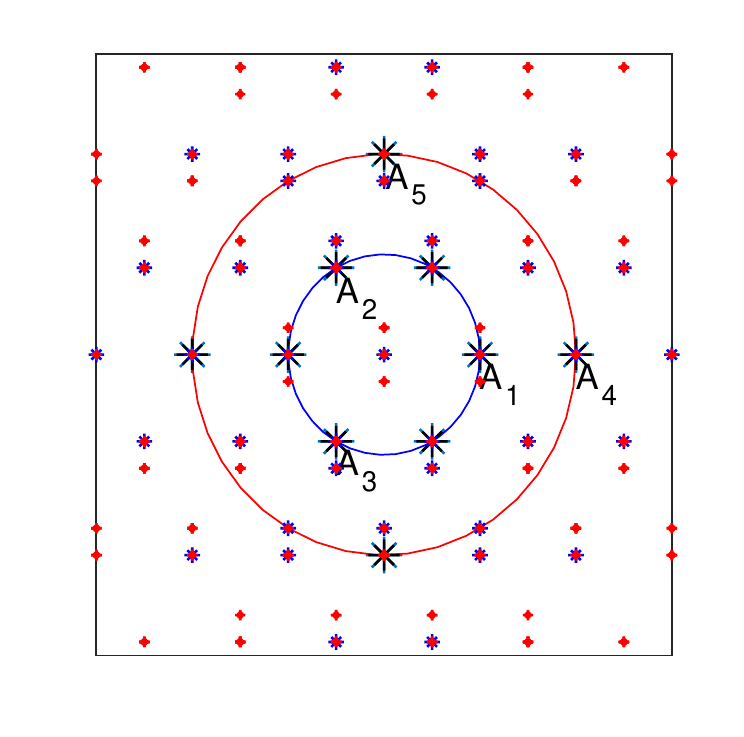}
\ece
\vs{-2mm}
   \caption{{\small Critical wave vectors (black stars), and first three 'layers' 
of the generated quasi-lattice. 
  \label{qcfig}}}
\end{figure}

In Cell 4 of {\tt cmdsqc} we set up a lattice as indicated in 
Fig.~\ref{qcfig}, and let \ass\ return the coefficients in 
the 1st and 4th equation to obtain 
\hugast{
{\tt Q}=\bpm  4& -1& 2c_2& 1& 1& c_2\\
-2& -3& 2c_2& 0& 0&  0
\epm \label{qqc}
}
and 
\hugast{\label{cqc} 
{\tt C}=\bpm 
1& 1& -1&        6c_2^2 + 3c_3& 1&  4& -1&    201c_2^2/25 + 6c_3\\
 1& 2& -2&       24c_2^2 + 6c_3& 1& -2& -3&          36c_2^2 + 6c_3\\
 1& 3& -3&       24c_2^2 + 6c_3& 2&  4& -2&   1636c_2^2/81 + 6c_3\\
 1& 4& -4& 201c_2^2/25 + 6c_3& 3&  4& -3&   1636c_2^2/81 + 6c_3\\
 1& 5& -5&       12c_2^2 + 6c_3& 4&  4& -4& 8102c_2^2/2025 + 3c_3\\
 2& 3&  4&       36c_2^2 + 6c_3& 4&  5& -5&    204c_2^2/49 + 6c_3
\epm. 
}
There are two quadratic terms mapping on $k_1=(1,0)$ 
but only one mapping to $k_4=(2,0)$, and therefore the second row 
of the second block of {\tt Q} is filled with zeros. 

\def\dhome{./KS}
\subsection{A damped Kuramoto-Sivashinsky type of equation, demo {\tt KS}}
\label{ksd}
The demo KS deals with damped/driven Kuramoto-Sivashinsky (KS) \cite{KY76,Siv77} 
type of equations, e.g., in 1D, 
\huga{\label{KS}
\pa_t u=-(1+\pa_x^2)^2 u+\lam u+c_2\pa_{x}(u^2), 
}
i.e., a SH equation with a convective nonlinearity, which gives 
another important class of pattern forming systems. Generalizing to $\R^d$, 
in Fourier space, $(\pa_{x_1}+\ldots+\pa_{x_d})(u^2)$ becomes 
$\hat{f}(u)=\hat{q}(k)\CF(u^2)(k)$, with the symbol 
$\hat{q}(k)=\ri(k_1+\ldots+k_d)$. To deal with such problems, 
\ass\ has the optional field {\tt p.qs}, where the user can put 
a function handle to the desired symbol for quadratic terms. For instance, 
for $\hat{q}(k)=\ri(k_1+\ldots+k_d)$ we define 
\fbox{{\tt function q=qs1(k); q=sum(k)*1i; end}}. 
Then calling {\tt p.sb=1; p.qs=@qs1; syms c2; p.c2=c2; p.c3=0; 
[Q,C]=ampsys(p);} 
(see script {\tt KS/cmdsKS.m}) 
in 1D we obtain {\tt Q=[]} (no quadratic interaction in 1D), and 
\hugast{
{\tt C}=(1,1,-1,-4c_2^2/9),\quad \text{ i.e. } 
\ddT A_1=A_1-\frac 4 9c_2^2|A_1|^2A_1,}
which can be quickly checked by hand. 

Another canonical form of the 
nonlinearity, in particular in higher space dimensions, is $c_2|\nabla u|^2$, 
where $|\nabla u|^2:=(\pa_{x_1} u)^2+\ldots+
(\pa_{x_d} u)^2$. In this case $\hat{q}(k)=-|k|^2$, and 
setting {\tt p.qs=@qs2} with 
\fbox{{\tt function q=qs2(k); q=-norm(k)$\hat{\ }$2; end}}  
and computing on a 3D FCC we obtain {\tt Q=[]} again, and 
\alinon{ {\tt C}=
  \begin{pmatrix}
    1 & 1 & -1 & 8c_2^2/9\\
    1 & 2 & -2 & 1296c_2^2/25\\
    1 & 3 & -3 & 1296c_2^2/25\\
    1 & 4 & -4 & 1296c_2^2/25\\
    -2 & -3 & -4 & 144c_2^2
  \end{pmatrix}. 
} 
As usual, the coefficients in the equations for $\ddT A_j$, $j=2,3,4$ 
can be obtained from symmetry, or {\tt ampsys} with suitable {\tt p.eqnr}.

%\lstinputlisting[caption={{\small {\tt KS/cmdsKS.m}.}},
%label=KSl,language=matlab,stepnumber=5,
%firstnumber=1]{\dhome/cmdsKS.m} 

%\input{brSC}

%\def\dhome{/hh/path/pde2path/demos/asdemos/brussel}
\def\dhome{./brussel}
\subsection{The Brusselator model, demo {\tt brussel}}\label{brd}
As an RD example we consider the Brusselator \reff{bruss1} 
({\tt asdemos/brussel}), over square (2D) and BCC (3D) 
lattices. Listing \ref{brf} shows the straightforward 
implementation of the 'nonlinearity' $f$,  which is independent 
of the lattice, while Listing \ref{brcmds} shows the commands 
for computing the AEs over different lattices, and some 
comparison to the analytic formulas \reff{cijana}. 

\lstinputlisting[caption={{\small {\tt brussel/f.m}, see comments for explanation.}},
label=brf,language=matlab,stepnumber=5,
firstnumber=1]{\dhome/f.m}

\begin{lstlisting}[caption={{\small {\tt brussel/cmdsBr.m}, script for 
\reff{bruss1} over various lattices.
%, organized in \mlab\ cells, i.e., run cell-by-cell.
}},
label=brcmds,language=matlab,stepnumber=5,linerange=1-29, 
firstnumber=1]
%% standard Brusselator, C1: Preparations, clear (possible) previous setups of p 
% and set parameters;  to be run as prep. for Cells 2-5 below
p=[];  
a=2; R=0.9; bc=(1+R)^2; b=bc; D1=1; D2=(a/R)^2; % parameters 
uh=[a; b/a]; kc=sqrt(R/D1); % derived data (u*, crit.wave number kc) 
p.par=[a,b]; p.uh=uh; p.D=[D1 0; 0 D2]; % put data into p 
p.bifpar=2; p.del=1e-3; bp=b+p.del; p.uhp=[a; bp/a]; % settings for mu'(b_c,k_c): 
% bifpar is b, p.uhp is u* at b+del, p.del is used for FD for mu' 
%% C2: 2D square lattice, with comparison to analytic formulas for lam and c31
p.k=wavevec(kc,21); [Q,C,c1]=ampsys(p);  % compute ampsys over squares 
t0=(a^2-R^2)/(a^2*(1+R)); lam=1/bc; 
c3a=(8-38*R-5*R^2+8*R^3)/(9*R*(a^2-R^2)); 
fprintf('(c1,c31,lam,c3a)=(%g, %g, %g, %g)\n', c1,C(1,end),lam/t0,c3a); 
%% C3: 3D BCC, with p.cons=1, yielding c32=c33=c34=2*c31 
p.cons=1; p.k=wavevec(kc,33); [Q,C,c1]=ampsys(p); % BCC 
q=Q(1,end); c31=C(1,end); fprintf('(c1,q,c31)=(%g %g %g)\n', c1,q,c31); 
t0=(a^2-R^2)/(a^2*(1+R)); lam=1/bc; c1a=lam/t0; 
qa=2*a*(1-R)/(a^2-R^2); c3a=(8-38*R-5*R^2+8*R^3)/(9*R*(a^2-R^2));
fprintf('(c1a,qa,c3a)=(%g %g %g)\n', c1a,qa,c3a); % analytic formulas: 
%% C4: BCC, with cons=0, yielding c31 as in formula, and different c32,c33,c34
p.cons=0; p.k=wavevec(kc,33); [Q,C,c1]=ampsys(p); % BCC 
q=Q(1,end); c31=C(1,end); c32=C(2,end); c33=C(4,end); c34=C(7,end); 
fprintf('(c3a1,c3a)=(%g %g)\n', c31,c3a); 
fprintf('(c1,q,c31,c32,c33,c34)=(%g %g %g %g %g %g)\n', c1,q,c31,c32,c33,c34); 
%% C5: same as C5 with R=0.75, showing more sign.deviations from c33=2*c31 etc
R=0.4; bc=(1+R)^2; b=bc; D1=1; D2=(a/R)^2; % update parameters 
uh=[a;b/a];kc=sqrt(R/D1);p.par=[a,b]; p.uh=uh; p.D=[D1 0;0 D2]; % put data into p 
bp=b+p.del; p.uhp=[a; bp/a]; p.k=wavevec(kc,33);
p.cons=0; [Q,C,c1]=ampsys(p); % BCC 
\end{lstlisting} 

The first cell in Listing \ref{brcmds} contains preparations for 
subsequent calls to \ass. Here, the idea of the cell mode 
is that the user can run (just) the first cell to set/change parameters, 
and then choose the lattice and call \ass\ for the new parameters. 
This is our typical operational mode. Thus, in lines 5 and 6 we set up the 
parameters as needed, namely the problem parameters $a,b$ (using the auxiliary parameter 
$R$ to set $b$ and $D_2$) including the diffusion coefficients, and the 
homogeneous steady state $U^*$. In line 7 we put this data into {\tt p}, 
and in line 8 we provide the data needed to compute 
$\mu'(k_c)=\pa_b \mu(k_c)$. We set {\tt p.bifpar=2} as we take $b$ as 
the bifurcation parameter. 

\subsubsection{Squares.}\label{brd2d}
In Cell 2 of Listing \ref{brcmds} we set a 2D square lattice, 
and call \ass\ to compute the coefficients $c_1, c_{31}, 
c_{32}$ as in \reff{tim}. The output is {\tt Q=[]} (no quadratic 
resonances on the square lattice), {\tt c1=0.66}, and 
\huga{\label{csq}
{\tt C}=\bpm 1&1&-1&-0.945\\
1&2&-2&-1.462\epm,
}
i.e. $c_{31}=-0.945$ and $c_{32}=-1.462$, in the notation from \reff{tim}. 
This agrees with \reff{cijana}, and this also holds for other values 
$a,R$ chosen in line 5. 

\subsubsection{The BCC lattice}\label{brd3d}
In Cell 3 of Listing \ref{brcmds} we consider the BCC lattice, where 
by symmetry the general form of the first amplitude equation is \reff{bccgen}, 
now with $\lam=c_1(B-B_c)$. 
For $c_1$ and $c_{31}$ (with {\tt p.cons=0}) 
we naturally have the formulas \reff{cijana} again, 
and $a_{12}$ has been computed in \cite{VWDB92} to 
\huga{\label{lamcoeff} 
a_{12}=\frac{4A(1{-}R)}{A^2{-}R^2}, 
}
which shows that $|R-1|$ should be small. Moreover, in the 
limit $R\to 1$ we obtain 
\huga{\label{c3jlim}
\text{$c_{32}/c_{31}\to 2, c_{33}/c_{31}\to 2, c_{34}/c_{31}\to 2$ as $
R\to 1$,}
}
and $c_{31}\to -1$ for the choice $A=2$, fixed in Cell 1. For {\tt p.cons=1} 
we always have 
\huga{\label{cjsimp}
\text{$c_{32}=c_{33}=c_{34}=2c_{31}$ also for $R\ne 1$.}} 
% which holds for {\tt p.cons=1}, \reff{bccgen} can be simplified to 
% {\small 
% \huga{\label{aef}
% \begin{aligned}
% f_1= &\lambda A_1{+}\frac{a}2(A_2 \ov{A}_6{+}A_3A_5){+}b(0.5|A_1|^2{+}|A_2|^2{+}|A_3|^2{+}|A_4|^2{+}|A_5|^2{+}|A_6|^2)A_1{+}b(A_2A_4A_5{+}A_3\ov{A}_4\ov{A}_6), 
% \end{aligned}
% }
% } 
% with $\lam=c_1(B-B_c)$, $a=a_{12}$, and $b=2c_{31}$,  
% and the other $f_j$ obtained from symmetry. 
% The amplitude system in this form has been succesfully 
% used in \cite{UW18b} 
% for predictions about BCC patterns $\ubcc$, tubes $\utub$, 
% and steady fronts between $\ubcc$ and $u=u^*$, and between $\ubcc$ 
% and $\utub$. 

Using \ass, we can check the formulas \reff{lamcoeff} for 
$c_1, a_{12}$ and $c_{31}$, and compute $c_{32},\ldots,c_{34}$ 
to check \reff{c3jlim}, respectively quantify the deviations from 
the limit. In C3 we run \ass\ with {\tt p.cons=1}. 
The output is $c_1=0.66$, ${\tt Q}=\bpm 2&-6&0.125\\
3& 5&0.125\epm$ and 
$$
{\tt C}=\bpm 1&1&-1&-0.846\\
1&2&-2&-1.693\\
1&3&-3&-1.693\\
\vdots&\vdots&\vdots&\vdots\\
3&-4&-6&-1.693
\epm,
$$
yielding the correct values for $c_1, a_{12}$, and \reff{cjsimp} 
holds, 
but the value for $c_{31}$ naturally differs from $c_{31}\approx -0.945$ 
from \reff{cijana}. 
On the other hand, using {\tt p.cons=0} in C4 yields $c_1,Q$ as before, and 
$$
{\tt C}=\bpm 1&1&-1&-0.945\\
1&2&-2&-1.999\\
1&3&-3&-1.999\\
1&4&-4&-1.462\\
1&5&-5&-1.999\\
1&6&-6&-1.999\\
2&4&5&-2.041\\
3&-4&-6&-2.041
\epm. 
$$
Thus, $c_{31}$ agrees with \reff{cijana}, but there is 
a deviation from \reff{cjsimp} in particular 
for $c_{33}$ (4th row of {\tt C}). This gets worse for 
smaller $R$ in C5, as expected. 

\def\dhome{./ExtBrus}
\subsection{An extended Brusselator as a three component system, 
demo {\tt ExtBrus}.}
\label{ebrusd} 
As a 3-component example in \ass\ consider the extended Brusselator \cite{Ep02}
\begin{equation}\label{exbr}
  \begin{aligned}
    &\dot{u}_1=D_1\Delta u_1+a-(1+b)u_1+u_1^2u_2-cu_1+du_3,\\
    &\dot{u}_2=D_2\Delta u_2+bu_1-u_1^2u_2,\\
    &\dot{u}_3=D_3\Delta u_3+cu_1-du_3,\\
  \end{aligned}
\end{equation}
which also serves as a tutorial example 
in \cite{hotheo}, see also \cite{hotutb}. 
A homogeneous steady state is given by $(u_1,u_2,u_3)=(a,b/a,ac/d)$, and 
for fixed $(D_1,D_2,D_3)=(0.01,0.1,1)$ this undergoes Hopf--, wave-- 
or Turing bifurcations as the parameters $(a,b,c,d)$ vary. 
For convenience, in {\tt ExtBrus} we include a short script 
{\tt plotev} to plot the dispersion relation. 
Fixing $(a,c,d)=(1.08,1,1)$, the first instability is a Turing 
instability at $b_c\approx 3.057$, with $k_c\approx 6.83$. 
Listing \ref{ebrcmds} shows the script file to compute the Landau 
coefficients for 2 simple cases, giving, e.g., $c_1=0.893$, $Q=[]$ and 
$C=\bpm 1&1&-1&-1.098\\
1&2&-2&92.877\epm$ on the square. 

\begin{lstlisting}[caption={{\small {\tt ExtBrus/cmds.m}, script for 
\reff{exbr} in 1D, and over 2D square lattice.}},
label=ebrcmds,language=matlab,stepnumber=5,linerange=1-7, 
firstnumber=1]
%% preparations 
p=[]; Dx=0.01; Dy=0.1; Dz=1; p.D=[Dx 0 0;  0 Dy 0;  0 0 Dz]; 
a=1.08; b=3.057; p.par=[a b]; p.uh=[a; b/a; a]; 
p.del=del; p.uhp=[a; (b+p.del)/a; a]; p.bifpar=2; kc=6.83; 
%% calling ampsys, 1D, and 2D square
p.k=wavevec(kc,1); [Q,C,c1,phi]=ampsys(p); pause 
p.k=wavevec(kc,21); [Q,C,c1,phi]=ampsys(p); 
\end{lstlisting} 

%\bibliography{bibi}{}
%\bibliography{/home/hu/hubib}\bibliographystyle{alpha}
\newcommand{\etalchar}[1]{$^{#1}$}

\end{document}